\documentclass[12pt]{article}
\usepackage{graphicx}
\usepackage{amsmath}
\usepackage{cite}
\usepackage[normalem]{ulem}
\usepackage{color}
\def\beq{\begin{equation}}
\def\eeq{\end{equation}}
\def\bea{\begin{eqnarray}}
\def\eea{\end{eqnarray}}
\def\nn{\nonumber}

\def\roughly#1{\mathrel{\raise.3ex\hbox
{$#1$\kern-.75em\lower1ex\hbox{$\sim$}}}}

\def\bd{B^0_d}
\def\bdbar{{\bar B}^0_d}
\def\bs{B^0_s}

\newcommand{\A}{\mathcal{A}}
\newcommand{\e}{\epsilon}

\newcommand{\Ab}{\bar{\mathcal{A}}}
\newcommand{\Abf}{\bar{\mathcal{A}_f}}
\def \BBbr{$B^0$-${\bar B}^0$\ }
\def \BBbrd{$\bd$-$\bdbar$ \ }
\def \oln{\overline}
\def\T {${T}$\ }
\def\CP {${C\!P}$\ }
\def\CPT {${C\!P T}$\ }

\def\bra#1{\langle #1 \vert}
\def\ket#1{\vert #1 \rangle}
\def\braket#1{\langle #1 \rangle}
\def\eqref#1{(\ref{#1})}

\begin{document}

\begin{flushright}
IMSc/2017/12/10 \\ 
UdeM-GPP-TH-17-260 \\
\end{flushright}

\begin{center}
\bigskip
{\Large \bf \boldmath Using time-dependent indirect \CP asymmetries
  \\ to measure $T$ and \CPT violation in $B^0$-${\bar B}^0$ mixing} 
\\
\bigskip
\bigskip
{\large
Anirban Karan $^{a,}$\footnote{kanirban@imsc.res.in},
Abinash Kumar Nayak $^{a,}$\footnote{abinashkn@imsc.res.in},
Rahul Sinha $^{a,}$\footnote{sinha@imsc.res.in}, \\
and David London $^{a,}$\footnote{london@lps.umontreal.ca}
}
\end{center}

\begin{flushleft}
~~~~~~~~~~~$a$: {\it The Institute of Mathematical Sciences, HBNI,}\\
~~~~~~~~~~~~~~~{\it Taramani, Chennai 600113, India,}\\
~~~~~~~~~~~$b$: {\it Physique des Particules, Universit\'e de Montr\'eal,}\\
~~~~~~~~~~~~~~~{\it C.P. 6128, succ. centre-ville, Montr\'eal, QC, Canada H3C 3J7}
\end{flushleft}

\begin{center}
\bigskip (\today)
\vskip0.5cm {\Large Abstract\\} \vskip3truemm
\parbox[t]{\textwidth}{Quantum field theory, which is the basis for
  all of particle physics, requires that all processes respect \CPT
  invariance. It is therefore of paramount importance to test the
  validity of \CPT conservation. In this Letter, we show that the
  time-dependent, indirect \CP asymmetries involving $B$ decays to a
  \CP eigenstate contain enough information to measure $T$ and \CPT
  violation in $B^0$-${\bar B}^0$ mixing, in addition to the standard
  $CP$-violating weak phases. Entangled $B^0{\bar B}^0$ states are not
  required (so that this analysis can be carried out at LHCb, as well
  as at the $B$ factories), penguin pollution need not be neglected,
  and the measurements can be made using $\bd$ or $\bs$ mesons.}

\end{center}

\thispagestyle{empty}
\newpage
\setcounter{page}{1}
% Decrease texheight (for preprint numbers) again
%\textheight 23.0 true cm
\baselineskip=14pt

\CPT invariance is one of the fundamental principles of quantum
field theory: all physical processes are expected to respect this
symmetry.  Indeed, \CPT violation would have a profound impact on
physics in general, as it would also lead to a violation of Lorentz
symmetry \cite{CPTLorentz1,CPTLorentz2}. Given its importance to the
theoretical framework underlying all of particle physics, much
attention has been devoted to experimentally testing the validity of
\CPT invariance.

One of the consequences of \CPT invariance in quantum field theory is
that a particle and its antiparticle should have the same mass and
lifetime. However, these quantities are mostly dominated by the strong
or electromagnetic interactions. Given that \CPT violation, if
nonzero, is certainly a very small effect, it is very difficult to
test it by measuring the differences of masses or lifetimes. A more
promising area for testing \CPT violation is in $P^0$-${\bar P}^0$
mixing, where $P^0$ is a neutral pseudoscalar meson (e.g., $K^0$,
$D^0$, $\bd$, $\bs$) \cite{Lavoura:1999ma}. Since this mixing is a
second-order electroweak process, small $CPT$-violating effects may be
easier to detect.

Now, it is known that, in addition to incorporating \CPT violation,
the most general \BBbr mixing matrix also involves $T$ and \CP
violation. As a consequence, studying \CPT violation is impossible
without discriminating it from the effects of \CP and $T$ violation.
That is, the effects of $CP$, $T$ and \CPT violation must be considered
together. 

A first step was taken in Refs.~\cite{CPTVBmix_th1, CPTVBmix_th2},
with followup papers in Refs.~\cite{Alvarez:2003kh, Alvarez:2004tj,
  CPTVBmix_th3, CPTVBmix_th4, CPTVBmix_th5, Bernabeu:2016sgz}. The
proposed method uses entangled $B^0{\bar B}^0$ states produced in the
decay of the $\Upsilon(4S)$, with one meson decaying to a CP
eigenstate ($J/\psi K_S$ or $J/\psi K_L$) and the other used to tag
the flavour. Using this technique, true $T$- and $CPT$-violating
asymmetries can be measured. The BaBar Collaboration implemented this
strategy \cite{CPTVBmix_BaBar1, CPTVBmix_BaBar2}, culminating in the
measurement of $T$ violation \cite{Lees:2012uka}.

At present, all experimental results are consistent with \CPT
conservation. On the other hand, an important improvement in
statistics is expected at LHCb and Belle II, so that it will be
possible to measure the $CP$-, $T$- and $CPT$-violating parameters
with greater precision. However, the method using entangled states
produced in the decay of the $\Upsilon(4S)$ cannot be used at LHCb. An
alternate approach is needed.

In this Letter, we re-examine the possibilities for measuring $T$ and
\CPT violation in \BBbr mixing using the decays of $B^0$ or ${\bar
  B}^0$ to a \CP eigenstate.  As we will show, the time-dependent
indirect \CP asymmetry contains sufficient information to measure the
conventional $CP$-violating effects and extract the $T$- and
$CPT$-violating parameters \cite{CPTVstatus,LHCbmeas}. Since no true
$T$- and $CPT$-violating asymmetries are measured, this is an indirect
determination of the $T$- and $CPT$-violating parameters.  In this
sense, this method is complementary to that using entangled states.

We focus on $\bd$-$\bdbar$ mixing but the same approach can be
modified and applied to the $\bs$ system.  Note that we restrict the
analysis to $T$ and \CPT violation arising from the \BBbr mixing
matrix alone.  If there are new-physics contributions to $B$ decays,
we assume they are $CPT$-conserving.

We begin by reviewing the most general formalism for \BBbr mixing, in
which \CPT and \T violation are incorporated. The $2\times 2$
hermitian matrices $\mathbf{M}$ and $\mathbf{\Gamma}$, respectively
the mass and decay matrices, are defined in the $(B^0, {\bar B}^0)$
basis. When $\mathbf{M}-(i/2) \mathbf{\Gamma}$ is diagonalized, its
eigenstates are the physical light ($L$) and heavy ($H$) states $B_L$
and $B_H$.  Now, any $2\times2$ matrix can be expanded in terms of the
three Pauli matrices $\sigma_i$ and the unit matrix with complex
coefficients:
\beq
\label{eq:Eq1}
\mathbf{M} - \frac{i}{2} \mathbf{\Gamma} = E_1 \sigma_1 + E_2 \sigma_2 + E_3 \sigma_3 - i D\mathbf{1} ~.
\eeq
Comparing both sides of this equation, we obtain
\bea
\label{EDdefs}
E_1 &=& {\rm Re} \, M_{12}-\frac{i}{2}{\rm Re} \, \Gamma_{12} ~, \nn\\
E_2 &=& -{\rm Im} \, M_{12} + \frac{i}{2} {\rm Im} \, \Gamma_{12} ~, \nn\\
E_3 &=& \frac{1}{2} \, (M_{11} - M_{22}) - \frac{i}{4} (\Gamma_{11} - \Gamma_{22}) ~, \nn\\
D &=& \frac{i}{2} (M_{11} + M_{22}) + \frac{1}{4} (\Gamma_{11} + \Gamma_{22}) ~.
\eea
We can define complex numbers $E$, $\theta$ and $\phi$ as follows:
\bea
\label{paramdefs}
E & = & \sqrt{E_1^2 + E_2^2 + E_3^2} ~, \nn\\
E_1 & = & E \sin\theta\cos\phi ~, \nn\\
E_2 & = & E \sin\theta\sin\phi ~, \nn\\
E_3 & = & E \cos\theta ~. 
\eea
The eigenvalues of $\mathbf{M}-(i/2) \mathbf{\Gamma}$ are $E -iD$ and
$-E - iD$, with eigenstates
\bea
\label{MState}
\ket{B_L} &=& p_1 \ket{B^0} + q_1 \ket{{\bar B}^0} ~, \nn\\
\ket{B_H} &=& p_2 \ket{B^0} - q_2 \ket{{\bar B}^0} ~,
\eea
where $p_1=\cos{\frac{\theta}{2}}$,
$q_1=e^{i\phi}\sin{\frac{\theta}{2}}$, $p_2=\sin{\frac{\theta}{2}}$
and $q_2=e^{i\phi}\cos{\frac{\theta}{2}}$.

In Ref.~\cite{TDLee} (pgs.\ 349-358), T.D. Lee discusses the \CPT and
\T properties of $\mathbf{M}$ and $\mathbf{\Gamma}$. First, if \CPT
invariance holds, then, independently of \T symmetry,
\beq
\label{CPTgood}
M_{11} = M_{22} ~,~~ \Gamma_{11} = \Gamma_{22}
\Longrightarrow E_3 = 0 \Longrightarrow \theta = \frac{\pi}{2} ~.
\eeq
In addition, if \T invariance holds, then, independently of \CPT
symmetry,
\beq
\label{Tgood}
\frac{\Gamma_{12}^*}{\Gamma_{12}} = \frac{M_{12}^*}{M_{12}}
\Longrightarrow {\rm Im} (E_1 E_2^*) = 0 \Longrightarrow {\rm Im} \, \phi = 0 ~.
\eeq
From Eqs.~(\ref{EDdefs}) and (\ref{paramdefs}), we have
\beq
e^{i\phi} = \pm \sqrt{ \frac{ M_{12}^* - i \Gamma_{12}^* / 2 }{ M_{12} - i \Gamma_{12} / 2 }} ~.
\eeq
Thus, if \T is a good symmetry, the left-hand quantity is a pure
phase, and the modulus of the square root is one.  Note that it is
usually said that the absence of \CP violation implies $|e^{i\phi}| =
1$. However, strictly speaking, this is due to the absence of $T$
violation. The two reasons are equivalent only if \CPT is conserved.

Second, defining $\zeta \equiv \langle S | L \rangle = \langle B_L |
B_H \rangle$,
\beq
\label{BLBH}
\zeta ~{\rm is} 
\begin{cases}
{\hbox{real ~~~~~~~~~~~ if \CPT holds,}} \\ 
{\hbox{imaginary ~~~ if \T holds.}} 
\end{cases}
\eeq
Using Eq.~(\ref{MState}), we have
\bea
\braket{B_L|B_H} & = & \cos{\frac{\theta}{2}} \sin{\frac{\theta^*}{2}} 
                            - e^{i\phi}\sin{\frac{\theta}{2}} e^{-i\phi^*}\cos{\frac{\theta^*}{2}} \nn\\
& = & \frac12 \left[ (1 - e^{-2 {\rm Im}(\phi)} ) \sin\{{\rm Re}(\theta)\} 
- i \{1 + e^{-2 {\rm Im}(\phi)}\}\sinh\{{\rm Im}(\theta)\}\right] ~. 
\eea
Now, if \CPT is a good symmetry, then $\theta = \frac{\pi}{2}$
[Eq.~(\ref{CPTgood})], so that $\sinh\{{\rm Im}(\theta)\} = 0$ and
$\braket{B_L|B_H}$ is real. And if \T is a good symmetry, then ${\rm
  Im}(\phi) = 0$ [Eq.~(\ref{Tgood})], so that $(1 - e^{-2 {\rm
    Im}(\phi)} ) = 0$ and $\braket{B_L|B_H}$ is imaginary. With this,
we see that Eq.~(\ref{BLBH}) is verified. (Obviously, if both \CPT and
\T are good symmetries, $\braket{B_L|B_H} = 0$, i.e., the states are
orthogonal.)

It will be useful to define the complex $\theta$ and $\phi$ in terms
of real parameters as $\theta=\theta_1 + i\theta_2$ and $\phi=\phi_1 +
i\phi_2$. In the absence of both $T$ and \CPT violation in
$B^0$-${\bar B}^0$ mixing, the parameters take the following values:
\beq
\label{defs1}
\theta_1 = \frac{\pi}{2} ~,~~ \theta_2 = 0 ~,~~ \phi_1 = -2\beta^{mix} ~,~~ \phi_2 = 0 ~,
\eeq
where $\beta^{mix}$ is the weak phase describing \BBbr mixing. In the
standard model, $\beta^{mix}=\beta$ for the $\bd$ meson. Now, if $T$
and \CPT violation are present in the mixing, the parameters
$\theta_1$, $\theta_2$ and $\phi_2$ will deviate from these values. We
define $\epsilon_{1,2,3}$ via
\beq
\label{defs2}
\theta_1 \to  \frac{\pi}{2} + \e_1 ~,~~ \theta_2 \to  \e_2 ~,~~ \phi_2 \to \e_3 ~.
\eeq
$\e_1$ and $\e_2$ are $CPT$-violating parameters, whereas $\e_3$
indicates $T$ violation.

The values for $\e_1$, $\e_2$ and $\e_3$ have been reported by the
BaBar and Belle Collaborations \cite{CPTVBmix_Belle,
  Lees:2014kep}. Their notation is related to ours as follows:
\beq
\cos\theta \leftrightarrow  -z ~,~~ \sin\theta \leftrightarrow \sqrt{1-z^2} ~,~~ e^{i\phi} \leftrightarrow \frac{q}{p} ~,
\eeq
so that
\beq
\label{epsdefs}
\e_1 = {\rm Re}(z) ~,~~ \e_2 = {\rm Im}(z) ~,~~ \e_3 = 1-\Big|\frac{q}{p}\Big| ~. 
\eeq
$\e_1$ and $\e_2$ are expected to be very small, as they are
$CPT$-violating parameters. As for $\e_3$, note that $|q/p|$ has been
measured at the $\Upsilon(4S)$ using the same-sign dilepton asymmetry,
assuming \CPT conservation \cite{HFAG}:
\beq
\label{qoverp}
\Big|\frac{q}{p}\Big| = 1.0010 \pm 0.0008 ~~~ \Longrightarrow ~~~
\epsilon_3 = -(1.0 \pm 0.8) \times 10^{-3} ~.
\eeq
Thus, $\e_3$ is also very small.

Above, we have called $\e_1$ and $\e_2$ the $CPT$-violating
parameters. But one must be careful about such names. $\e_1$ and
$\e_2$ do not contribute {\it only} to observables measuring \CPT
violation. They also lead to $CP$- and $T$-violating effects.
Similarly, the $T$-violating parameter $\e_3$ also contributes to
$CP$-violating observables. And the reverse is true: recall that, in
Ref.~\cite{Lees:2012uka}, the BaBar Collaboration measured a large
true $T$-violating asymmetry. This does {\it not} suggest that $\e_3$
is large, as there are also large contributions to the asymmetry
coming from $CP$-violating effects (assuming \CPT invariance). The
point is that $\e_1$, $\e_2$ and $\e_3$ are also sources of $CP$
violation, and it is this fact that allows their measurement in the
time-dependent indirect \CP asymmetry, as we will see below.

In the presence of $T$ and \CPT violation in $B^0$-${\bar B}^0$
mixing, the time evolution of the flavor eigenstates ($\ket{B^0}
\equiv \ket{B^0}(t=0)$ and $\ket{{\bar B}^0} \equiv \ket{{\bar
    B}^0}(t=0)$) is given by
\bea
\label{TState}
\ket{B^0(t)} &=& ( g_{+} + g_{-}\cos{\theta})\ket{B^0}  + e^{i \phi} g_{-} \sin{\theta} \ket{{\bar B}^0} ~, \\
\ket{{\bar B}^0(t)} &=& e^{-i \phi} g_{-} \sin{\theta} \ket{B^0} + ( g_{+} - g_{-} \cos{\theta}) \ket{{\bar B}^0} ~, \nn
\eea
where $g_{\pm} = (e^{-iH_Lt} \pm e^{-iH_Ht})/2$:
\bea
g_+ &=& e^{-i(M-i \frac{\Gamma}{2})t} \cos\left(\left(\Delta{M}-i\frac{\Delta\Gamma }{2}\right)\frac{t}{2}\right) ~, \nn \\
g_- &=& e^{-i(M-i \frac{\Gamma}{2})t} i \sin\left(\left(\Delta{M}-i\frac{\Delta\Gamma }{2}\right)\frac{t}{2}\right) ~.
\eea
Here $M \equiv (M_H + M_L)/2$, $\Delta M \equiv M_H-M_L$, $\Gamma
\equiv (\Gamma_H + \Gamma_L)/2$ and $\Delta\Gamma \equiv
\Gamma_H-\Gamma_L$.

We consider a final state $f$ to which both $B^0$ and ${\bar B}^0$ can
decay. Using Eq.~\eqref{TState}, the time-dependent decay amplitudes
for uncorrelated or tagged neutral mesons are given by
\bea
%	\label{TAmplitude}
\mathcal{A}(B^0(t) \to f) &\!=\!& (g_{+} + g_{-} \cos{\theta} ) \mathcal{A}_{f} 
+ e^{i \phi} g_{-} \sin{\theta} \bar{\mathcal{A}}_{f} ~, \\
\mathcal{A}({\bar B}^0(t)\to  f) &\!=\!& e^{-i \phi} g_{-} \sin{\theta} 
	\mathcal{A}_{f} + (g_{+} - g_{-} \cos{\theta} )\bar{\mathcal{A}}_{f} ~, \nn 
\eea
where $\mathcal{A}_f \equiv \bra{f}\mathcal{H}_{\Delta F=1}\ket{B^0}$
and $\bar{\mathcal{A}}_{f} \equiv \bra{f}\mathcal{H}_{\Delta
  F=1}\ket{{\bar B}^0}$. The differential decay rates
$\Gamma_f(B^0(t)\to f)$ and $\Gamma_f({\bar B}^0(t)\to f)$ are given
by\footnote{In Ref.~\cite{Bernabeu:2016sgz} it was pointed out that
  the coefficient of $\cos(\Delta M t)$ includes a $CPT$-violating
  piece.}
\bea
	\label{TGamma}
		\frac{d\Gamma}{dt}(B^0(t)\to   f) &=& \frac12 e^{-\Gamma t} 
		\left[
\sinh\left(\Delta\Gamma t/2\right)  \left\{ 2\rm{Re} 
		\left( \cos\theta |\mathcal{A}_{f}|^2 +e^{i \phi} \sin\theta 
		\mathcal{A}_{f}^{*} \bar{\mathcal{A}}_{f} \right) \right\}
		\right. \nn \\
		&& \hskip-1truein  +~\cosh\left(\Delta\Gamma t/2\right) \left\{ |\mathcal{A}_f |^2 + |\cos\theta |^2|\mathcal{A}_f |^2 + |e^{i \phi}\sin\theta |^2|\bar{\mathcal{A}}_{f} |^2 
		+ 2 \rm{Re}\left(e^{i \phi} \cos\theta^{*} \sin\theta 
		\mathcal{A}_{f}^{*} \bar{\mathcal{A}}_{f} \right) \right\} \nn\\
		&& \hskip-1truein  +~\cos(\Delta M t) \left\{ |\mathcal{A}_f |^2 - |\cos\theta |^2|\mathcal{A}_f |^2 - |e^{i \phi}\sin\theta |^2|\bar{\mathcal{A}}_{f} |^2 - 2 \rm{Re}\left(e^{i \phi} \cos\theta^{*} \sin\theta 
		\mathcal{A}_{f}^{*} \bar{\mathcal{A}}_{f} \right)
		\right\} \nn\\
		&& \left. -~\sin(\Delta Mt) \left\{ 2 \rm{Im} \left( \cos\theta  |\mathcal{A}_{f}|^2 + e^{i \phi} \sin\theta \mathcal{A}_{f}^{*}\bar{\mathcal{A}}_{f}
		\right) \right\}
		\right] ~, \\
	\label{TGammaBar}
		\frac{d\Gamma}{dt}({\bar B}^0(t)\to  f) &=& \frac12 e^{-\Gamma t} \left[
		\sinh\left(\Delta\Gamma t/2\right) \left\{ 2\rm{Re} 
		\left( -\cos\theta^{*} |\bar{\mathcal{A}}_{f}|^2 + e^{i \phi^*} 
		\sin\theta^{*} \mathcal{A}_{f}^{*} \bar{\mathcal{A}}_{f} \right) \right\} \right. \nn \\
		&& \hskip-1truein +~\cosh\left(\Delta\Gamma t/2\right) \left\{ |\oln{\mathcal{A}_f} |^2 + |\cos\theta |^2|\oln{\mathcal{A}_f} |^2 + | e^{-i \phi}\sin\theta |^2|\mathcal{A}_{f} |^2 
		- 2 \rm{Re}\left( e^{i \phi^*} \cos\theta \sin\theta^{*} \mathcal{A}_{f}^{*} \bar{\mathcal{A}}_{f} \right)
		\right\} \nn\\
		&& \hskip-1truein +~\cos(\Delta M t) \left\{ |\oln{\mathcal{A}_f} |^2 - |\cos\theta |^2|\oln{\mathcal{A}_f} |^2 - | e^{-i \phi}\sin\theta |^2|\mathcal{A}_{f} |^2 + 2 \rm{Re}\left( e^{i \phi^*} \cos\theta \sin\theta^{*} 
		\mathcal{A}_{f}^{*} \bar{\mathcal{A}}_{f} \right)
		\right\} \nn\\
		&& \left. +~\sin(\Delta Mt) \left\{ 2 \rm{Im} \left( -\cos\theta^{*}  
		|\bar{\mathcal{A}}_{f}|^2 +  e^{i \phi^*} \sin\theta^{*} 
		\mathcal{A}_{f}^{*} \bar{\mathcal{A}}_{f}
		\right) \right\} \right] ~.
\eea
If we set $\theta = \frac{\pi}{2}$ and ${\rm Im} \, \phi = 0$ in the
above expressions, we recover expressions for the differential decay
rates that are commonly found elsewhere in the literature.

We now write $\theta=\theta_1 + i\theta_2$ and $\phi=\phi_1 +
i\phi_2$, with [see Eqs.~(\ref{defs1}) and (\ref{defs2})]
\beq
\theta_1 \to  \frac{\pi}{2} + \e_1 ~,~~ \theta_2 \to  \e_2 ~,~~ \phi_1 = -2\beta^{mix} ~,~~ \phi_2 \to \e_3 ~,
\eeq
where the $\e_{1,2,3}$ are very small.  In order to probe $T$ and \CPT
violation in \BBbr mixing, one must measure the parameters
$\epsilon_{1,2,3}$. Below, we illustrate how this can be done in the
$\bd$ system. For $\bs$ mesons, the procedure is similar, though a bit
more complicated.

First, as regards $\Gamma_d$, the value of $y_d =
\Delta\Gamma_d/2\Gamma_d$ has been measured to be small: $y_d =
-0.003\pm 0.015$ with the $\bd$ lifetime of $1.520\pm 0.004$ ps
\cite{pdg}. This means that we can approximate $\sinh (\Delta\Gamma
t/2) \simeq \Delta\Gamma t/2 = y_d \Gamma_d t$ and $\cosh
(\Delta\Gamma t/2) \simeq 1$. In principle, for large enough times,
this approximation will break down. However, even at time scales of
$\mathcal{O}(10)$ ps, the approximation holds to $\sim 10^{-4}$, and
by this time most of the $\bd$s will have decayed.

The observable we will use to extract the $T$- and $CPT$-violating
parameters $\epsilon_{1,2,3}$ is the time-dependent indirect \CP
asymmetry $A_{CP}^f(t)$ involving $B$-meson decays to a \CP
eigenstate. It is defined as
\beq
A_{CP}^f(t) =\! \frac{d\Gamma/dt({\bar B}_d^0(t)\to f_{CP})-d\Gamma/dt(B_d^0(t)\to f_{CP})}
                     {d\Gamma/dt({\bar B}_d^0(t)\to f_{CP})+d\Gamma/dt(B_d^0(t)\to f_{CP})} ~.
\eeq
In the limit of \CPT conservation, $T$ conservation in the mixing, and
$\Delta\Gamma = 0$, one has the familiar expression
\beq
A_{CP}^f(t) =\! S \sin(\Delta M_d t) - C \cos(\Delta M_d t),
\eeq
where
\bea
\varphi & \equiv & \phi_1 - \arg[\A_f] + \arg[\Abf] ~, \\ 
C & \equiv & \frac{|\A_f|^2 - |\Ab_f|^2}{|\A_f|^2 + |\Ab_f|^2 } ~, \\
S & \equiv & \sqrt{1-C^2} \ \sin\varphi ~.
\eea
Here, $C$ is the direct \CP asymmetry and $\varphi$ is the measured
weak phase, which differs from the mixing phase $\phi_1$ if
$\arg[\A_f]\neq \arg[\Abf]$. If there is no penguin pollution, then
$\varphi$ cleanly measures a weak phase and $C = 0$. But if there is
penguin pollution, then neither of these holds.

In the presence of $T$ and \CPT violation in the mixing, we use
Eqs.~\eqref{TGamma} and \eqref{TGammaBar} to obtain the time-dependent
\CP asymmetry. We first expand the various functions in the two
equations, keeping only terms at most linear in the small quantities
$\epsilon_{1,2,3}$ and $\Delta\Gamma_d$:
\bea
\label{numer}
\frac{d\Gamma}{dt}(\bar B_d^0(t)\to f) - \frac{d\Gamma}{dt}(B_d^0(t)\to   f) & = & e^{-\Gamma_d t} (|\A_f|^2 + |\Abf|^2) 
\left[ \e_3 + \sqrt{1-C^2} \e_1 \cos\varphi  \right. \\ 
		&& \hskip-6truecm \left. +~\cos(\Delta M_d t) \left\{ -C - \e_3 - \sqrt{1-C^2}\e_1\cos\varphi  \right\}
		+\sin(\Delta M_d t) \left\{ -\e_2 + \sqrt{1-C^2} \sin\varphi \right\} \right] ~, \nn \\
\label{denom}
\frac{d\Gamma}{dt}(\bar B_d^0(t)\to f) + \frac{d\Gamma}{dt}(B_d^0(t)\to   f) & = & e^{-\Gamma_d t} (|\A_f|^2 + |\Abf|^2) \nn\\
&& \hskip-6truecm \left[ 1 + C \e_3 + \frac12 \sqrt{1-C^2} \Delta\Gamma_d t \cos\varphi - \sqrt{1-C^2}\e_2\sin\varphi \right. \\ 
		&& \hskip-5.5truecm \left. +~\cos(\Delta M_d t) \left\{ - C \e_3 + \sqrt{1-C^2} \e_2\sin\varphi \right\}
		+\sin(\Delta M_d t) \left\{ C \e_2 + \sqrt{1-C^2}\e_3 \sin\varphi \right\} \right] ~. \nn
\eea
The denominator [Eq.~(\ref{denom})] has the form $A(1 + x)$, with $x$
small, so we can appriximate $1/A(1+x) \simeq (1 - x)/A$. Combining all
the pieces, and again keeping only terms at most linear in
$\epsilon_{1,2,3}$ and $y_d$, we obtain\footnote{A time-dependent \CP
  asymmetry having a complicated form with higher harmonics in $\Delta
  M_d t$, similar to that in Eq.~\eqref{ACPpsNP}, was noted in
  Ref.~\cite{CPTVBmix_th4}.}
\bea
\label{ACPpsNP}
A_{CP/CPT}^f(t) & \simeq & c_0 + c_1 \cos(\Delta M_d t) + c_2 \cos(2\Delta M_d t) + s_1 \sin(\Delta M_d t) + s_2 \sin(2\Delta M_d t) \nn\\
&& +~c_1^\prime\, \Gamma_d\,t\cos(\Delta M_d t)+s_1^\prime\, \Gamma_d\,t\sin(\Delta M_d t) ~,
\eea
where the coefficients are given by
\bea
\label{ACPpsNP2}
c_0 & = & \e_1 \cos\varphi + \e_3  -\frac{1}{2} \e_3 \sin^2\varphi ~, \nn \\
c_1 & = & -C-\e_3-\e_1 \cos\varphi-\e_2 C\sin\varphi ~, \nn \\
c_2 & = & \frac{1}{2}\e_3\sin^2\varphi +\e_2 C \sin\varphi ~, \nn \\
s_1 & = & \sqrt{1-C^2}\sin\varphi-\e_2\cos^2\varphi -\e_3C\sin\varphi ~, \nn\\
s_2 & = & -\frac{1}{2} \e_2\sin^2\varphi+\e_3 C\sin\varphi ~, \nn \\
c_1^\prime & = & C\,y_d^{}\,\cos\varphi ~, \nn\\
s_1^\prime & = & -\frac{1}{2}\,y_d^{}\,\sin 2\varphi ~.
\eea
The seven pieces have different time dependences so that, by fitting
$A_{CP/CPT}^f(t)$ to the seven time-dependent functions, all
coefficients can be extracted.

The five observables $c_0$, $c_1$, $c_2$, $s_1$ and $s_2$ can be used
to solve for the five unknown parameters $C$, $\varphi$ and
$\epsilon_{1,2,3}$. In practice, a fit will probably be used, but
there is an analytical solution. The parameter $C$ is simply 
given by
\beq
\label{NCPsol-1}
C = -(c_0+c_1+c_2) ~.
\eeq
The solution for $\sin\varphi$ is obtained by solving the following
quartic equation:
\bea
\label{NCPsol-2}
&& \sin^4\varphi-2\bigg[\frac{s_1+2s_2}{2-C^2}\bigg] \sin^3\varphi 
+4C\bigg[C+\frac{c_2}{2-C^2}\bigg]\sin^2\varphi\nn\\
&& \hskip2truecm -~4\bigg[\frac{2C^2(s_1+s_2)-s_2}{2-C^2}\bigg]\sin\varphi-\bigg[\frac{8C\, 
	c_2}{2-C^2}\bigg]=0 ~.
\eea
Of course, there are four solutions, but, since the $\epsilon_i$ are
small, the correct solution is the one that is roughly
$s_1/\sqrt{1-C^2}$.  Finally, $\e_1,\e_2,\e_3$ are given by
\bea
\label{NCPsol-3}
\epsilon_1&=&c_0\sec\varphi-\frac{(2-\sin^2\varphi)(c_2 \sin\varphi+2C\, 
	s_2)}{(4C^2+\sin^2\varphi)\,\sin\varphi\,\cos\varphi} ~,\nn\\
\epsilon_2&=&\frac{2\,(2C\,c_2-s_2\,\sin\varphi)}{(4C^2+\sin^2\varphi)\,\sin\varphi} ~,\nn\\
\epsilon_3&=&\frac{2\,(c_2\,\sin\varphi+2C\,s_2)}{(4C^2+\sin^2\varphi)\,\sin\varphi} ~.
\eea
This shows that it is possible to measure the parameters describing
$T$ and \CPT violation in \BBbrd mixing using the time-dependent
indirect \CP asymmetry, and this can be carried out at LHCb.

The parameters $c'_1$ and $s'_1$ depend only on $\varphi$ and
$y_d$. Thus, given knowledge of $\varphi$, the value of $y_d$ can be
found from measurements of these parameters. Note that, even if the
width difference $\Delta\Gamma_d$ between the two $B$-meson
eigenstates vanishes, the $T$-violating parameter $\epsilon_3$ can
still be extracted. This is contrary to the claim of
Refs.~\cite{Alvarez:2003kh} and \cite{CPTVBmix_th4}.

Above, the method was described for the $\bd$ system. In the case of
$\bs$ mesons, $\Delta\Gamma_s$ is not that small, so the functions
$\sinh\left(\Delta\Gamma_s t/2\right)$ and $\cosh\left(\Delta\Gamma_s
t/2\right)$ must be kept throughout. This modifies the forms of
Eqs.~(\ref{numer}), (\ref{denom}) and (\ref{ACPpsNP}), but the idea
does not change. $A_{CP/CPT}^f(t)$ still depends on seven different
time-dependent functions, a fit can be performed to extract their
coefficients, and $C$, $\varphi$, $\epsilon_{1,2,3}$ and
$\Delta\Gamma_S$ can be found using the measurements of these
coefficients.

Finally, we have another handle for probing \CPT violation in \BBbrd
mixing. At present, we know that $\epsilon_3 = -(1.0 \pm 0.8) \times
10^{-3}$ [Eq.~(\ref{qoverp})]. Now, suppose that there is no \CPT
violation (i.e., $\epsilon_1 =\epsilon_2 = 0$). In this case, for the
time-dependent \CP asymmetry [Eq.~(\ref{ACPpsNP})], we can eliminate
$C$ and $\sin\varphi$ using Eqs.~\eqref{NCPsol-1}-\eqref{NCPsol-3}.
The coefficients $c_0, c_2$ and $s_2$ can then be expressed in terms
of the measured quantities $c_1$, $s_1$ and $\epsilon_3$ as follows:
\bea
\label{cSol}
c_0 &=& \epsilon_3 \bigg[1-\frac{2 s_1^2}{(2-c_1^2+\epsilon_3^2)^2}\bigg] ~, \nn\\
c_2 &=& \frac{2 s_1^2\, \epsilon_3}{(2-c_1^2+\epsilon_3^2)^2} ~, \nn\\
s_2&=&-\frac{2 s_1\,(c_1+\epsilon_3)\, \epsilon_3}{(2-c_1^2+\epsilon_3^2)} ~.
\eea
The values of $c_1$ and $s_1$ have been measured for several $\bd$
decays to \CP eigenstates \cite{HFAG}, and the value of $\e_3$ is
independent of the decay mode. Using these values, we can estimate
$c_0$, $c_2$ and $s_2$ from Eq.~\eqref{cSol}, which assumes that \CPT
is conserved. As an example, for the final state $J/\psi K_S$, we find
\bea
c_0 & = & (-15.18 \pm 15.50) \times 10^{-4} ~, \nn\\
c_2 & = & (-4.31 \pm 4.41) \times 10^{-4} ~, \nn\\
s_2 & = & (0.29 \pm 0.43) \times 10^{-4} ~.
\eea
Should the measurements of $c_0$, $c_2$ and $s_2$ deviate
significantly from the above values, this would indicate the presence
of \CPT violation in \BBbrd mixing.

To sum up, we have shown that the time-dependent, indirect \CP
asymmetries involving $B^0,{\bar B}^0 \to f_{CP}$ contain enough
information to extract not only the $CP$-violating weak phases, but
also the parameters describing $T$ and \CPT violation in $B^0$-${\bar
  B}^0$ mixing. These measurements can be made at the $\Upsilon(4S)$
(e.g., BaBar, Belle) or at high energies (e.g., LHCb).  There is no
need to neglect penguin pollution in the decay, and the method can be
applied to $\bd$- or $\bs$-meson decays.

\bigskip
\noindent
{\bf Acknowledgments}: This work was financially supported in part by
NSERC of Canada (DL).

\end{document}